\documentclass[prd,showpacs,nofootinbib,floatfix,preprintnumbers]{revtex4}%

\usepackage{amssymb,hyperref,amsmath}
\usepackage[dvips]{color}
\usepackage[dvips]{graphicx}
\usepackage{epsfig}
\preprint{JLAB-THY-09-985}

\usepackage{latexsym} 

\begin{document}

\newcommand{\beq}{\begin{equation}}
\newcommand{\eeq}{\end{equation}}

\bibliographystyle{apsrev}

\title{A novel quark-field creation operator construction for hadronic physics
  in lattice QCD}

\author{Michael Peardon}
\email{mjp@maths.tcd.ie}
\affiliation{School of Mathematics, Trinity College, Dublin 2, Ireland}

\author{John Bulava}
\author{Justin Foley}
\author{Colin Morningstar}
\affiliation{Department of Physics, Carnegie Mellon University, Pittsburgh, PA
15213, USA}

\author{Jozef Dudek}
\author{Robert G. Edwards}
\author{B\'alint Jo\'o}
\author{Huey-Wen Lin}
\author{David G. Richards}
\affiliation{Thomas Jefferson National Accelerator Facility, Newport News, VA
23606, USA}

\author{Keisuke Jimmy Juge}
\affiliation{Department of Physics, University of the Pacific
Stockton, CA 95211, USA}

\author{for the {\em Hadron Spectrum Collaboration}}

\date{\today}


\begin{abstract}
A new quark-field smearing algorithm is defined which enables efficient
calculations of a broad range of hadron correlation functions.
The technique applies a low-rank operator to define smooth fields that are
to be used in hadron creation operators.
The resulting space of smooth fields is small enough that all
elements of the
reduced quark propagator can be computed exactly at reasonable computational
cost. Correlations between arbitrary sources, including multi-hadron
operators can be computed
{\em a posteriori} without requiring new lattice Dirac operator inversions.
The method is tested on realistic lattice sizes with light dynamical quarks.
\end{abstract}

\maketitle

\section{Introduction}
One of the goals of lattice calculations is to predict the low-energy hadron
spectrum of confined quarks and gluons, starting solely from the QCD 
lagrangian. This approach to spectroscopy necessitates methods for measuring the
two-point correlation functions of field operators with the selected quantum
numbers under investigation. A complete understanding of the QCD spectrum 
must also included excitations of the mesons and baryons as well as exotic
states. 
Describing the resonances seen in scattering experiments 
as states in QCD has long presented a challenge to lattice calculations as
direct access to the matrix elements related to decay widths is usually
missing in Euclidean formulations of field theory. The way to circumvent this
is well known in principle: decay properties can be inferred from a detailed
study of the dependence of the spectrum in a finite box on the volume
of the box \cite{DeWitt:1956be,Luscher:1991cf}.  
The best means of making solid determinations of energy levels is to employ a 
large basis of operators and then use the variational method to find good 
approximations to energy eigenstates. As soon as states above threshold are to 
be explored, the basis should include operators that resemble multi-hadron 
systems where the constituents have well-defined momenta. 
Access to all elements of the quark propagator \cite{Foley:2005ac} relevant 
to long-range correlation enables these measurements and 
also enables more detailed investigations into physical observables
whose accurate measurement has often eluded lattice practitioners. One example
is the isoscalar meson sector, where mass determinations require the
evaluation of disconnected diagrams as part of the two-point correlation
function.  As well as 
this ambitious list of requirements, precision data 
will be crucial for these calculations \cite{Morningstar:2008mc}.

In a Monte Carlo calculation on the lattice, the physically relevant signal in 
a correlation function falls exponentially and is rapidly dominated by
statistical fluctuations. Operators that create low-lying energy eigenstates
quickly are invaluable and improve the quality of data extracted 
exponentially. The most useful tool at the lattice practitioner's disposal in 
building good creation operators is smearing. The best means of
understanding the low-energy degrees of freedom of confinement must thus come 
from a smearing method that can address all these features; it must be 
numerically accessible, do a good job of projecting onto the lowest energy 
states and it must facilitate easy evaluation of correlation functions 
involving a broad set of creation operators, including those exciting 
multi-hadron states. 

A new generation of experiments devoted to hadron spectroscopy, 
including GlueX at Jefferson Lab, PANDA at GSI/FAIR and BES III
intend to make measurements with unprecedented precision and in
previously unexplored mass ranges and quantum-number sectors. The main aim of
the {\em Hadron Spectrum Collaboration} is to extract from realistic
lattice simulations predictions for masses, decay properties and relevant matrix
elements in these domains.

Spectroscopy investigations are almost as old as non-perturbative Monte Carlo
calculations on the lattice \cite{Hamber:1981zn,Weingarten:1980hx}.  As the
first studies progressed, it was quickly realised that operators
creating hadrons which are constructed directly from the fields in the lattice
lagrangian have significant overlap with a large tower of states and that
extracting the lightest elements of the spectrum was difficult. The solution was
to build operators from ``smeared'' fields, where some linear operator was
applied first to the quark degrees of freedom on the appropriate time-slice
before the creation operator was formed. The aim of constructing a smeared
field was to reduce the component of the fields close to the 
cut-off substantially, since these modes do not contribute significantly to 
long-range correlation functions.
Initially, the techniques considered smeared fields that were not
gauge covariant. Gauge-covariant schemes were introduced soon after
\cite{Gusken:1989ad,
Allton:1993wc}. In particular, Ref.~\cite{Allton:1993wc} introduced the Jacobi
smearing algorithm, in which a lattice approximation to the three-dimensional
gauge-covariant laplacian is constructed and applied iteratively to the field.
This acts naturally as a long-wavelength filter on the modes constituting the
quark field.

In this paper, a new method for smearing quark fields is described and tested in
large-scale realistic simulations.
In Sec.~\ref{sec:theory}, the formulation of the new algorithm is presented, 
including detail on its application to measurements of hadron two-point
correlation functions and three-point functions where an arbitrary current is
inserted.
Sec.~\ref{sec:results} presents the results of initial tests of the
effectiveness of the method, and Sec.~\ref{sec:discussion} discusses practical
issues that arise with the method and outlines some future research directions.

\section{Operator construction \label{sec:theory}}
The energy of an eigenstate of the hamiltonian of a quantum field 
theory can be determined by computing the correlation function between 
creation and annihilation operators $\chi$ at Euclidean times $t$ and $t'$;
\begin{equation}
   C(t',t)  = \Big\langle \chi(t')\,  \chi^\dagger(t) \Big\rangle.
\end{equation}
Inserting a complete set of eigenstates of the hamiltonian, such that 
$\hat{H} | k \rangle = E_k  | k\rangle$, this correlation function
decomposes into a sum of contributions from all states in the spectrum
with the same quantum numbers as the source operators,
\begin{equation}
   C(t',t)  = \sum_k \big|\langle \chi | k \rangle\big|^2 \, e^{-{E_k} (t'-t)}.
\end{equation}
In order to
measure energies of low-lying states, it it crucial to construct operators that
overlap predominantly with these light modes. 
This exposes the asymptotic behaviour of the correlator at earlier
time-separations and enables more statistically accurate determinations of
energies.
There is a good deal of freedom in
the construction of appropriate operators. Once the constraints of symmetries
and temporal localisation are imposed, any function of the fields in the path
integral can be used in principle.

Smearing is a well-established means of defining an initial step in the
construction of creation operators. Rather than applying a creation operator to
the fields directly, a smoothing function is applied first. The function should
preserve as many symmetries as possible while effectively removing the
presence of short-range modes, which make an insignificant contribution to the
low-energy correlation function. In contemporary simulations, a popular form
of this operation in theories where fermion fields couple to gauge bosons is
the Jacobi smearing method \cite{Allton:1993wc}.

Gauge-covariant quark smearings based on the lattice Laplacian start 
with the simplest representation of the second-order three-dimensional
differential operator
\begin{equation}
   -\nabla^2_{xy}(t) = 6 \delta_{xy} - \sum_{j=1}^3
      \left(
	  \tilde{U}_j(x,t)         \delta_{x+\hat\jmath,y}
	 +\tilde{U}^\dagger_j(x-\hat\jmath,t) \delta_{x-\hat\jmath,y}
	  \right),
\end{equation}
where the gauge fields, $\tilde{U}$ may be constructed from an appropriate
covariant gauge-field-smearing algorithm \cite{Morningstar:2003gk}. 
After defining the Laplace operator, a simple smearing can be written 
\begin{equation}
  J_{\sigma, n_\sigma}(t) = \left( 1 + \frac{\sigma \nabla^2(t)}{n_\sigma} \right)^{n_\sigma},
  \label{eqn:smearing}
\end{equation}
where $\sigma$ and $n_\sigma$ are tunable parameters that can be used to
optimise projection onto the states under investigation.
For large $n_\sigma$, this approximates the exponential of $\sigma\nabla^2$,
{\it i.e.}
\begin{equation}
  \lim_{n_\sigma\rightarrow\infty} J_{\sigma, n_\sigma}(t) = \exp\left(\sigma
      \nabla^2(t)\right).
\label{eqn:jacobi}
\end{equation}
The resulting exponential suppression of higher eigenmodes of the lattice 
Laplace operator means only a small number of the lowest modes contribute 
substantially to $J$.

This observation suggests the smearing operator can be approximated
by forming an eigenvector representation, truncated to the lowest
modes.  Since the smearing operators used in lattice calculations are
approximated well by low-rank constructions, a definition of smearing can be
chosen to enforce this more absolutely.  Let $V_M$ be the vector space
of scalar fields charged under the fundamental representation of the gauge
group on a particular time-slice. $V_M$ has rank $M = N_c \times N_x \times
N_y \times N_z$ where $N_c$ is the number of colours, and $N_x,N_y$ and $N_z$
are the extents of the lattice in the three spatial directions.
Now define smearing to be a well-chosen operator of rank $N \ll M$.  This class
of operators will be called ``distillation'' operators.  There is a substantial
benefit to doing this: if the rank of the operator is sufficiently small, all
elements of the propagation matrix from this space can be constructed at an
affordable computational cost (which is proportional to the rank of the space, $N$).
Consequently, correlation functions involving arbitrarily intricate hadron
creation operators can be measured with a fixed inversion overhead.

Define the distillation operator on time-slice $t$ as a product of an $M \times
N$ matrix and its hermitian conjugate:
\begin{equation}
 \Box(t) = V(t) V^\dagger(t)
\implies \Box_{xy}(t) = \sum_{k=1}^N v_x^{(k)} (t) v_y^{(k)\dag} (t). 
    \label{eqn:box}
\end{equation}
The $k^{\rm th}$ column of $V(t)$ contains the $k^{\rm th}$ eigenvector
of $\nabla^2$ evaluated on the background of the spatial gauge fields of
time-slice $t$, once the eigenvectors have been sorted by eigenvalue. This is
the projection operator into $V_N$, the subspace spanned by these eigenmodes,
so $\Box^2 = \Box$.
When the number of eigenvectors included is the same as the dimension of
$V_M$, {\em i.e.} $N = M$, the distillation operator becomes the identity, and
fields acted upon are unsmeared.

The Laplace operator inherits many symmetries of the vacuum. It transforms
like a scalar under rotations, is covariant under gauge transformation and
is parity and charge conjugation invariant. If the action of one of these
symmetries on the Laplace operator maps $\nabla^2 $ onto $\tilde\nabla^2$, then
there is a unitary transformation, $R$ on $V_M$ such that
\beq
R \tilde\nabla^2 R^\dagger = \nabla^2.
\eeq
This implies that if $v$ is an eigenvector of $\nabla^2$ then the eigenvectors
of $\tilde\nabla^2$ are $R v$.
Considering the definition of the distillation operator given in
Eq.~\ref{eqn:box}, the transformed operator must then obey
\beq
R \stackrel{\sim}{\Box} R^\dagger = \Box, 
\eeq 
and so correlation functions constructed using distilled fields have the same
symmetry properties on the lattice as those constructed using Laplacian smearing
methods. 

  \subsection{Meson two-point correlation functions}\label{subsec:theory-mesons}
  Consider the momentum-projected creation and annihilation
operators of an isovector meson, $\bar{u} \Gamma^A d$ and
$\bar{d} \Gamma^B u$, where $\Gamma$ acts in spin and color as well as
coordinate space. Applying the distillation operator $\Box$ onto each
quark field, the creation operator at three-momentum $\vec{p}$ is written as
\begin{equation}
\chi^\dagger_M(\vec{p},t) 
  = \bar{u}_x(t) \Box_{xy}(t) \cdot e^{-i p\cdot y}\, \Gamma^A_{yz}(t) \cdot \Box_{zw}(t) d_w(t),
\label{eqn:meson_quark_op}
\end{equation}
where there is an implied summation over repeated spatial indices. In a shorthand notation the correlation function can be written as
\begin{equation}
  C^{(2)}_M(t',t) = \Big\langle \bar{d}(t')\Box(t')\Gamma^B(t')\Box(t')u(t')\,\cdot\,
                      \bar{u}(t)\Box(t)\Gamma^A(t)\Box(t) d(t)
    \Big\rangle.
\end{equation}
After evaluating the quark field path-integral and inserting the outer-product 
definition of the distillation operator $\Box$ from Eq.~\ref{eqn:box}, the 
correlator can be written
\begin{equation}
  C^{(2)}_M(t',t) = \mbox{Tr} \Bigl[
      \Phi^B(t') \,
      \tau(t',t) \,\Phi^A(t) \, \tau(t,t') \Bigr],
\end{equation}
where
\begin{equation}
  \Phi^A_{\alpha\beta}(t) = V^\dagger(t) \left[ \Gamma^A(t) \right]_{\alpha\beta} V(t)
  \equiv V^\dagger(t) {\cal D}^A(t) V(t) S^A_{\alpha\beta} ,
\label{eqn:meson_op}
\end{equation}
and
\begin{equation}
  \tau_{\alpha\beta }(t',t) = V^\dagger(t') M^{-1}_{\alpha\beta}(t',t) V(t),
\label{eqn:peram}
\end{equation}
with $M$ the lattice representation of the Dirac operator 
and where the quark spin indices, $\alpha,\beta$ of $\Phi$ and $\tau$ have been
explicitly written. $\Phi$ has a well-defined momentum, while there is no
explicit momentum projection in the definition of $\tau$. Often, $\Phi$ can be
decomposed into terms that act only within coordinate and color space ${\cal
D}^A$ and only within spin space $S^{A}$. Note that $\Phi$ and $\tau$ are
square matrices of dimension $N \times N_\sigma$ where $N_\sigma$ is the number
of components in a lattice Dirac spinor. Therefore it requires just
$N \times N_\sigma$ operations of the inverse of the fermion matrix on a
vector in order to compute all elements of $\tau$, the ``perambulator''.
Notice also that the choice of source and sink operators is entirely
independent of the computation of $\tau$; any source and sink operators can be
correlated {\it a posteriori} once all elements of the $\tau$ matrix have been
computed and stored. The method straightforwardly extends to the determination
of correlation functions for mesons composed of different, non-degenerate
quark flavors.

In the determination of an isoscalar meson correlation function, evaluation of
disconnected terms is required.
The disconnected diagram can be similarly represented once distilled fields are
used in creation and annihilation operators. Such a term would
comprise two separate traces over the distillation space:
\begin{equation}
  C^{(2,\mathrm{disc})}_M(t',t) =
   \mbox{Tr}\Bigl[ \Phi^A(t) \tau(t,t) \Bigr]
   \mbox{Tr}\Bigl[ \Phi^B(t') \tau^\dagger(t' ,t') \Bigr].
\end{equation}
An exact determination of this expression requires computing $\tau(t,t)$ for 
all time-slices $t$. 

Since distilled single-particle operators can be projected onto definite
momentum at both source and sink, multi-meson correlators can also be
constructed using creation and annihilation operators of the form
\begin{equation}
\chi_{MM}(\vec{q}=\vec{p}_1 - \vec{p}_2; t) = \chi_M(\vec{p}_1,t)
  \chi_M(-\vec{p}_2,t),
\label{eqn:multi_meson_quark_op}
\end{equation}
where $p_1$ and $p_2$ are the three-momenta of the single particle operators
in Eq.~\ref{eqn:meson_quark_op}. After integration over the quark fields, the 
resulting diagrams can again be computed by taking traces over products of
construction operators in the distillation space with the perambulators.
The precise structure of these functions depends on the quark flavor content of
the source and sink operators, so details are not presented here.  As before,
these matrices are small compared to the dimension of the space of quark
fields, and once the quark propagation is encoded in the perambulator,
these multi-hadron correlation function can be computed.

  \subsection{Baryon two-point correlation functions}
  The factorization technique can also be applied to baryons.
To illustrate the concepts involved, consider just the isospin-1/2 sector
although the technique generalizes naturally to other baryon multiplets. 
An annihilation operator involving displacements
as well as coefficients in spin space is written:
\begin{equation}
\chi_B(t) = \epsilon^{abc} S_{\alpha_1\alpha_2\alpha_3}
({\cal D}_1\Box d)^a_{\alpha_1}
({\cal D}_2\Box u)^b_{\alpha_2}
({\cal D}_3\Box u)^c_{\alpha_3}(t),
\end{equation}
where the color indices of the quark fields acted upon
by the displacement operators ${\cal D}_i$ are contracted with the
antisymmetric tensor, and repeated spin indices are summed.
After integration over quark fields, the correlation function
\begin{equation}
   C^{(2)}_B(t',t) = -\Big\langle \chi_B(t') \, \bar{\chi}_B(t)\Big\rangle
\end{equation}
can be factored into perambulator terms, Eq.~\ref{eqn:peram}
as well as creation and annihilation operators, where the latter can be 
written as
\begin{equation}
 \Phi^{(i,j,k)}_{\alpha_1\alpha_2\alpha_3}(t) = \epsilon^{abc}
\left({\cal D}_1 v^{(i)}\right)^{a}
\left({\cal D}_2 v^{(j)}\right)^{b}
\left({\cal D}_3 v^{(k)}\right)^{c}(t)\;
S_{\alpha_1\alpha_2\alpha_3}\ .
\label{eqn:baryon_op}
\end{equation}
The spin terms factor from the antisymmetric contraction of the vectors
$v$. 

There are two terms in the resulting correlation function
that involve tensor contractions of the creation and annihilation operators
with the perambulators
\begin{align}
C^{(2)}_B[\tau_d,\tau_u,\tau_u](t',t) &=
  \Phi^{(i,j,k)}(t')
  \tau_d^{(i,\bar{i})}(t',t)
  \tau_u^{(j,\bar{j})}(t',t)
  \tau_u^{(k,\bar{k})}(t',t)
  \Phi^{(\bar{i},\bar{j},\bar{k})*}(t)
  \nonumber\\
 &\quad- \Phi^{(i,j,k)}(t')
  \tau_d^{(i,\bar{i})}(t',t)
  \tau_u^{(j,\bar{k})}(t',t)
  \tau_u^{(k,\bar{j})}(t',t)
  \Phi^{(\bar{i},\bar{j},\bar{k})*}(t),
\end{align}
where there is an implicit tensor contraction over the internal spin
indices, and where the quark labels $d$ and $u$ are used to denote
the corresponding flavors of the quark perambulator terms in 
Eq.~\ref{eqn:peram}.

Similar to the meson case, the choice of source and sink operators is 
independent of the computation of the perambulators $\tau_d$ and $\tau_u$. 
The computation of these matrices can also be shared with computation of meson 
correlators.  A baryon correlation function can be evaluated {\em a posteriori} 
using the contractions of the vectors with displacements in 
Eq.~\ref{eqn:baryon_op} which can also be shared among the source and sink 
operators. These contractions do not involve spin components, thus making the 
storage of $\Phi$ manageable.
\label{subsec:theory-baryons}
  \subsection{Meson three-point correlation functions}\label{subsec:theory-mesons-3pt}
  A generic meson three-point function is written
\beq
   C^{(3)}(t_f, t, t_i) = \Big\langle
\big( \bar{d}\Box \Gamma^B \Box u\big)(t_f) \cdot
\big(\bar{u}\Gamma u\big)(t) \cdot
\big(\bar{u} \Box \Gamma^A \Box d \big)(t_i)
\Big\rangle,
\eeq
where there is no smearing in the operator on timeslice $t$. 
The completely connected Wick contraction can be expressed as
\beq
 C^{(3)}_{\mathrm{conn}}(t_f, t, t_i) = \mbox{Tr}\Bigl[\Phi^B(t_f)\, J(t_f,t,t_i) \,\Phi^A(t_i)\, \gamma_5 \tau^\dag(t_f,t_i) \gamma_5\Bigr]
\eeq
where the ``generalized perambulator'' is defined by
\begin{equation}
J_{\alpha\beta}(t_f,t,t_i) =  V^\dag(t_f) \left( M^{-1}(t_f,t)\, \Gamma(t) \,
M^{-1}(t,t_i)  \right)_{\alpha\beta} V(t_i). \label{eqn:genperam}
\end{equation}
In general, a disconnected term may also appear; this can be factorised as the
product of a two-point function (evaluated according to the algorithm of
Sec.~\ref{subsec:theory-mesons}) and a disconnected insertion. Since the
disconnected trace does not involve distilled fields, the evaluation of this 
insertion is not discussed further at this point. 
To compute the connected three-point correlation function, there are two 
distinct sets of inversions needed. The first is from the distillation vectors 
at the source time-slice $t_i$ and the second from the sink time-slice $t_f$. 
The generalised perambulator of Eq.~\ref{eqn:genperam} is then constructed by
contracting the two solutions from these two sets of inversions. 
The current insertion is encoded in the choice of operator $\Gamma$. 
Any momentum insertion at the current operator involves a Fourier transform on 
each time-slice $t$.

  \subsection{Baryon three-point correlation functions}\label{subsec:theory-baryons-3pt}
  Baryon three-point correlation functions can also be expressed in terms of 
the generalized perambulators of Eq.~\ref{eqn:genperam}.
Consider a three-point correlation function where a baryon is created on
time-slice $t_i$, then is acted on by a current operator on time-slice $t$ and
is subsequently annihilated at $t_f$:
\begin{equation}
   C^{(3)}_B(t_f, t, t_i) = - \Big\langle B(t_f)\cdot \big(\bar{u}\Gamma u \big)(t) \cdot \bar{B}(t_i) \Big\rangle.
\end{equation}
After quark-field integration, this can be written as a sum of both 
a connected three-point contribution
and the product of a disconnected insertion and a two-point correlator. 
As before, the discussion here is restricted to isospin-1/2 light baryons
although the technique is quite general. 
Note that the disconnected insertion involves unsmeared fields and must be 
evaluated through some other technique. The two-point contribution follows 
from Sec.~\ref{subsec:theory-baryons}.
For an up-quark insertion in the correlation function
\begin{equation}
\begin{split}
   C^{(3)}_\mathrm{conn.}(t_f, t, t_i) =& -\epsilon^{abc}\epsilon^{\bar{a}\bar{b}\bar{c}} S_{\alpha_1\alpha_2 \alpha_3} \bar{S}_{\bar{\alpha}_1 \bar{\alpha}_2 \bar{\alpha}_3}\\
&\quad\quad\Big\langle \big(
({\cal D}_1\Box d)^a_{\alpha_1}
({\cal D}_2\Box u)^b_{\alpha_2}
({\cal D}_3\Box u)^c_{\alpha_3} \big)(t_f)\\&\quad\quad\cdot
\big(\bar{u}\Gamma u\big)(t)
\cdot \big( (\bar d\Box\bar{\cal D}_1)^{\bar{a}}_{\bar\alpha_1}
(\bar u\Box\bar{\cal D}_2)^{\bar{b}}_{\bar\alpha_2}
(\bar u\Box\bar{\cal D}_3)^{\bar{c}}_{\bar\alpha_3}\big)(t_i)
\Big\rangle,
\end{split}
\end{equation}
there are four Wick contractions after quark-field integration.
The up-quark generalized perambulator from Eq.~\ref{eqn:genperam}
is denoted as
\begin{equation}
\widetilde{\tau}_U(t_f,t,t_i)  = 
V^\dagger(t_f) M_u^{-1}(t_f,t)\Gamma(t) M_u^{-1}(t,t_i) V(t_i) 
\end{equation}
and similarly for the down quark. Since the quarks in the insertion
come in pairs, the connected three-point correlator can be written as the 
sum of two two-point correlation functions 
with each up-quark perambulator substituted with the corresponding
generalized perambulator in turn;
\begin{equation}
C^{(3)}_u(t_f,t,t_i) = C^{(2)}[\tau_D,\tau_U,\widetilde{\tau}_U](t) +
C^{(2)}[\tau_D,\widetilde{\tau}_U,\tau_U](t).
\end{equation}
The down-quark contribution can be written as
\begin{equation}
C^{(3)}_d(t_f,t,t_i) = C^{(2)}[\widetilde{\tau}_D,\tau_U,\tau_U](t).
\end{equation}
This kind of construction is a generic feature of the three-point functions.

As in the previous two- and three-point correlator examples, the choice
of source and sink operators is independent of the perambulator and
generalized perambulator computations.  These latter generalized
perambulators can be shared among meson three-point correlator
calculations. In addition, for degenerate quarks, only one generalized
perambulator need be computed, independent of the flavor of the quark line 
where the current insertion is to occur. 
Thus, the overall computation of distinct three-point
correlation functions has a manageable computational cost.

\section{Test results\label{sec:results}}

To test the efficacy of the method, a first study of meson and baryon two-point
correlators was
performed. In addition, the method is demonstrated for a two-meson system.
The background gauge fields were generated by the {\em Hadron
Spectrum Collaboration} \cite{Edwards:2008ja,Lin:2008pr}. They are $16^3 \times 128$
and $20^3 \times 128$ anisotropic lattices, with the ratio of spatial to
temporal lattice spacings $a_s/a_t=3.5$ and 2+1 flavors of dynamical quarks.
The bare light-quark mass is $a_t m_q=-0.0840$, and when the $\Omega$ baryon
mass is used to set the scale, the corresponding pion mass is $383$ MeV.
The valence-quark action is the same as the light-dynamical-quark
action.

The numerical techniques for the iterative solution of the sparse linear system
for the Sheikholeslami-Wohlert improved discretisation of the Dirac equation
are described in Ref.~\cite{Edwards:2008ja}.  In particular, the EigCG
inverter~\cite{Stathopoulos:2007zi} was used for the solution of these linear
systems. 
This method uses eigenvector deflation to accelerate convergence.
The first few inversion calls are needed to construct a sufficient basis, and
subsequent inversions are accelerated by a factor of about three or more
compared to the undeflated solver. For the lattices considered here, this 
asymptotic rate is found for $N\ge 4$.

\label{subsec:results-intro}

  \subsection{The profile of the distillation operator}
      \label{subsec:results-wavefn}
  The distillation operator is written as a low-rank projection operator into a
vector space of smooth fields. As such, there is little physical intuition to
suggest it resembles more traditional smearing algorithms in generating a field
that is roughly localized in some confining region while still remaining 
smooth. To examine the properties of the operator, its spatial distribution 
was computed. Defining 
\beq
   \Psi(r) = \sum_{x,t} \sqrt{\mbox{Tr } \left(\Box_{x,x+r}(t) \Box_{x+r,x}(t)
       \right)}, 
\eeq
means $\Psi$ is a gauge-invariant measure of the degree to which a field is 
smeared by the application of the distillation operator. 
\begin{figure}[h]
\includegraphics[width=0.6\textwidth]{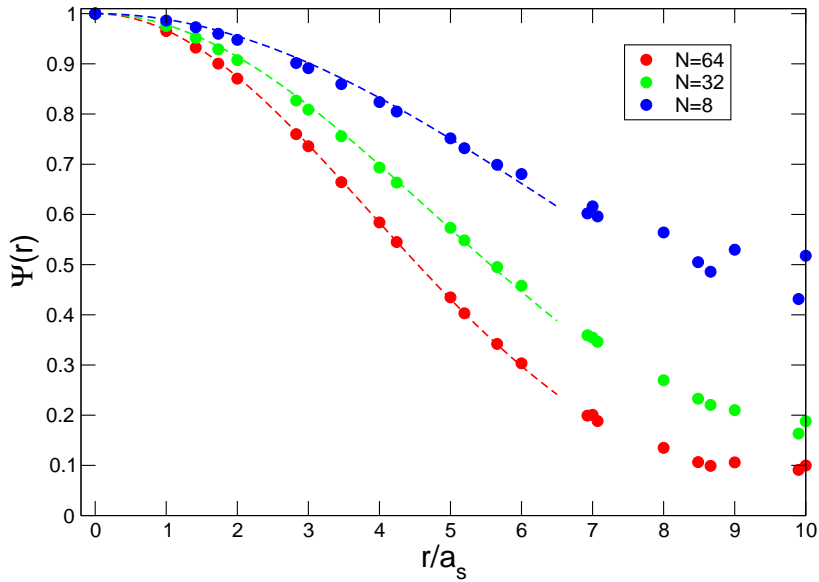}
\caption{\label{fig:wavefn} The spatial distribution of the distillation
  operator constructed from the lowest eigenvalues of the Laplace operator on
    the lattice. Data is from a $20^3$ spatial volume and statistical
    fluctuations are much smaller than plot symbols. The dashed lines are 
    fits to a gaussian distribution.} 
\end{figure}
This observable was measured for various values of $N$ and for a range of
different off-set vectors $r$ on a small ensemble of the $20^3$ spatial 
lattices and the results are 
presented in Figure~\ref{fig:wavefn}. The data clearly shows that for $N$ 
between 8 and 64 the distillation operator is very 
well described by a gaussian wavefunction. As $N$ increases, so the radius of
this distribution reduces. This behaviour is to be anticipated since when $N=M$
the distillation operator is the identity and has no spatial extent at all. 
The measurements presented in Fig.~\ref{fig:wavefn} also clearly illustrate 
the distillation operator has good rotational symmetry, indicating it 
is only very weakly influenced by dynamics at the lattice cut-off. The artifacts
seen at larger values of $r$ in the very broad $N=8$ distribution arise solely 
from the finite lattice volume used in these measurements. 

  \subsection{Meson correlation functions}\label{subsec:results-mesons}
\begin{figure*}
 \includegraphics[width=16cm]{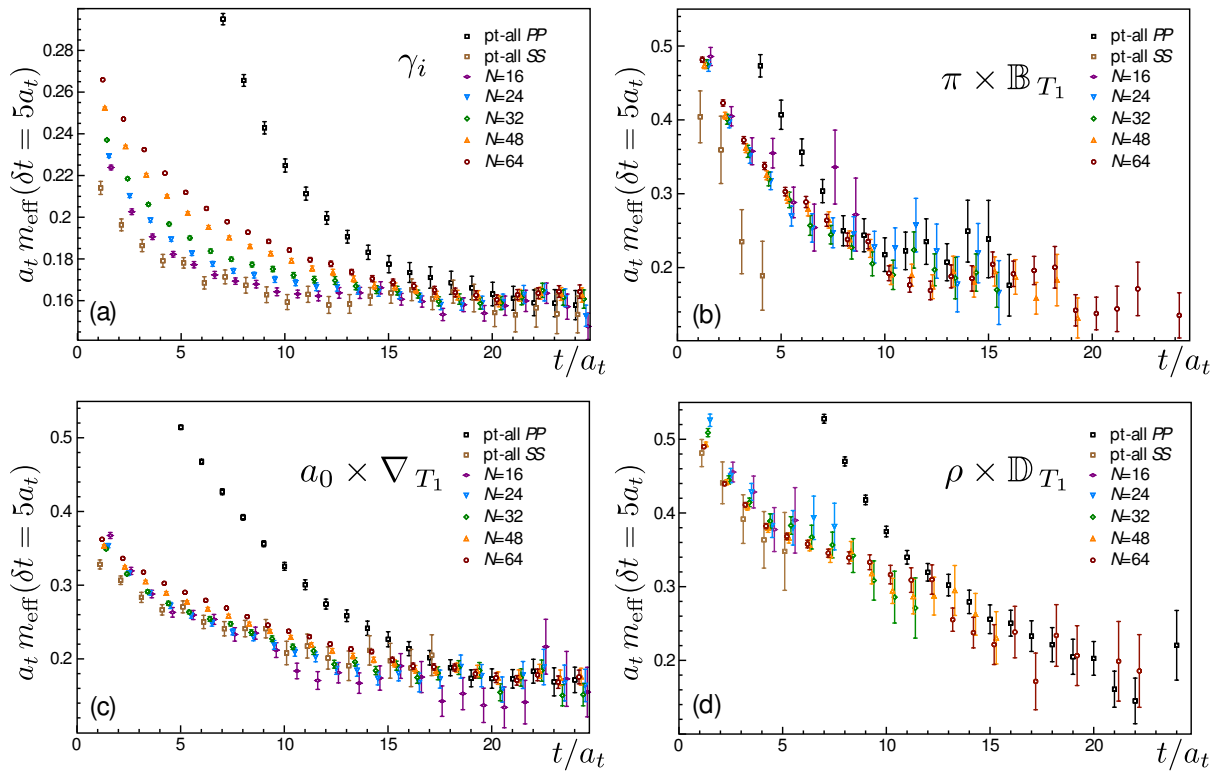}
\caption{\label{meff} Effective masses (five timeslice shift) for diagonal 
  correlators in the
$T_1^{--}$ channel computed on $16^3\times 128$ lattices. The operator
construction is shown in table \protect \ref{optab}. Correlators are
constructed using distilled quark fields with $N \in \{16,24,32,48,64\}$ vectors
and using point-to-all methods with standard gauge-invariant quark smearing 
(SS, brown) and without (PP, black). }
\end{figure*}

  In this section, connected meson correlation functions formed using the 
derivative-based operators presented in Refs.~\cite{Liao:2002rj,Dudek:2007wv} 
are shown, with particular focus on the $T_1^{--}$ channel.
Fig~\ref{meff} shows plots of the effective mass, defined by
\begin{equation}
m_{\mathrm{eff}}(t) = - \frac{1}{\delta t} \ln \left(\frac{C(t+\delta t)}{C(t)}\right)
\end{equation}
with $\delta t = 5 a_t$, for four diagonal correlators while varying the
number of distillation vectors. Also shown are the corresponding effective
masses computed using the traditional point-to-all propagator construction
with both unsmeared and Jacobi-smeared sources. The four operators used are
defined in Table \ref{optab}. In all cases, gauge links in the operators are
stout-smeared~\cite{Morningstar:2003gk}. 90 configurations of the
$16^3 \times 128$ lattice were used for this comparison.
\begin{table}
 \begin{tabular}{c | c}
Name & Construction\\
\hline
$\gamma_i$ & $\bar{\psi}\gamma_i\psi$ \\
$a_0\times \nabla_{T_1}$ & $ \bar{\psi} \nabla_i \psi$ \\
$\rho \times \mathbb{D}_{T_1}$ & $ |\epsilon_{ijk}|\, |\epsilon_{klm}|\,  \bar{\psi} \gamma_j  \nabla_l \nabla_m\psi$\\
$\pi \times \mathbb{B}_{T_1}$ & $ \epsilon_{ijk} \, \bar{\psi} \gamma_5 \nabla_j \nabla_k \psi$
\end{tabular}
\caption{\label{optab}Derivative-based meson operators used in these tests.
  $\vec{\nabla}$ is the one-link discretised covariant derivative}
\end{table}

A variational analysis \cite{Michael:1985ne,Luscher:1990ck} of matrices of
two-point correlation functions has proven to be a powerful method in
extracting the spectrum and matrix elements of excited states
\cite{Dudek:2006ej}. Consequently, cost estimates in terms of the computing 
time required to form such a matrix of correlators and not just any single
correlator should be considered.
As a measure of the relative computational demands of distillation versus the
traditional point-to-all approach, note that in the point-to-all approach,
ten forward propagators, generated by applying 
    $1$, $\nabla_{i=x,y,z}$, $\nabla_i \nabla_j|_{i \neq j}$ to a
    point source at the origin are required to produce the set of $T_1^{--}$
correlation functions defined above.  Each of these sources then requires twelve
inversions of the Dirac matrix ($N_c \times N_\sigma$).
To compare with the distillation methods, note that approximately the same
total inversion cost arises from using 32 distillation vectors since
\begin{equation}
(N_{\mathrm{srcs}}=10) \times (N_c = 3) \times (N_\sigma = 4) = 120 \approx (N = 32) \times (N_\sigma = 4) = 128.
\end{equation}

In the correlation function for the $\bar\psi\gamma_i\psi$ operator, it is
clearly seen that increasing the number of vectors causes the effective mass 
to approach that of the unsmeared point-to-all correlation function, since
$\sum_{k=1}^M v^{(k)} v^{(k)\dag} = 1$.
The statistical variance is observed to decrease rapidly as the number
of vectors is increased. For this range of $N$ it is
always smaller than the point-to-all correlators. A similar but less rapid
convergence of the signal is observed in the $a_0\times \nabla_{T_1}$ channel.

On the other hand, the $\pi \times \mathbb{B}_{T_1}$ signal appears to be
essentially unchanged over this range of vector number but the noise reduces
considerably with increasing vector number. The $\rho \times \mathbb{D}_{T_1}$ channel shows similar behaviour.
Note that these operators sample non-trivial spatial
dependence, and this may require a larger space of eigenvectors.
The rapid scaling of noise with number of vectors can be quantified as
follows: consider the correlator noise-to-signal ratio $r$ on a timeslice as a
function of the number of vectors used, $N$ and model this behaviour with
\begin{equation}
r = a + \frac{b}{N^p}. \label{noise}
\end{equation}
Fitting the parameters to best reproduce the Monte Carlo data from 450
configurations and averaging
over many time-slices gives the following estimates for
the scaling exponent
\begin{eqnarray}
p(\gamma_i) = 0.8(1); \quad p(a_0\times \nabla_{T_1}) = 1.1(2); \quad
p(\pi \times \mathbb{B}_{T_1}) = 1.3(3); \quad p(\rho\times \mathbb{D}_{T_1}) = 1.8(2). \nonumber
\end{eqnarray}
In all cases, the noise reduction is better than the behaviour expected from
simple statistical scaling, $p=0.5$.

\begin{figure}[h]
 \includegraphics[width=12cm]{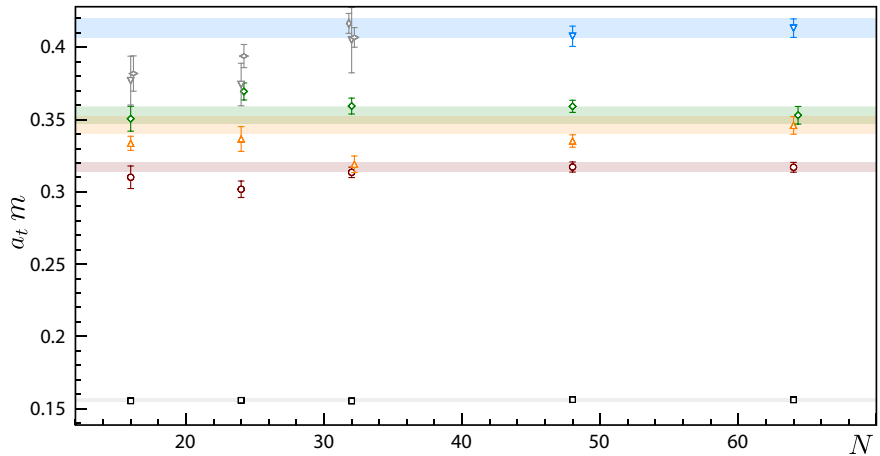}
\caption{\label{variational} $T_1^{--}$ variational mass spectrum as a
function of number of eigenvectors used. The
error bars are purely statistical and do not take account of possible 
systematic errors due to fitting.   }
\end{figure}

As a realistic test of the efficacy of distillation in the extraction
of an excited-state spectrum, a matrix of correlation
functions using twelve operators with $T_1^{--}$ quantum numbers was
constructed. In Fig.~\ref{variational} the low-lying spectrum extracted as a
function of number of vectors used is shown.
Clearly, 
even with a relatively small number of vectors this method 
can be used to construct two-point correlation matrices from which the
spectrum of low-lying states can be extracted with some precision. 
The masses, up through the fourth excited state, become more consistent
with increasing number of vectors.

\begin{figure*}
 \includegraphics[width=16cm]{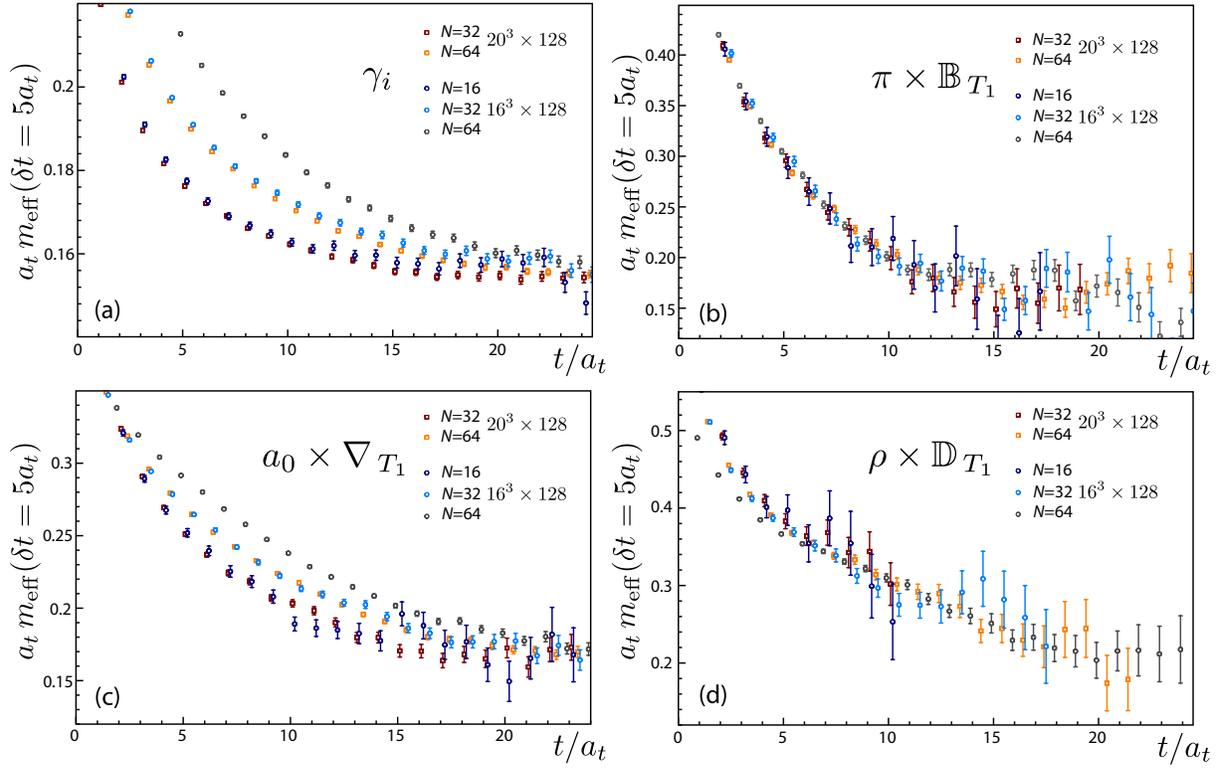}
\caption{\label{voltest} Effective masses (five timeslice shift) for diagonal correlators in the
$T_1^{--}$ channel on $16^3$ and $20^3$ spatial volumes. Operator construction shown in table
\protect \ref{optab}. }
\end{figure*}

Now consider the issue of the scaling of this method with increasing lattice
volume. Fig.~\ref{voltest} shows the effective masses of correlators computed
on both $16^3$ and $20^3$ lattices, the spatial-volume ratio here being very
close to 2 ($\tfrac{20^3}{16^3} = 1.95$). It is evident that the
correlation function with $N$ vectors on the $16^3$ volume is essentially the 
same as the corresponding case with $2N$ vectors on the $20^3$ lattice. Note 
that although the number of inversions for the same signal is doubled when the 
lattice volume increases from $16^3$ to $20^3$, there is a reduction in the 
noise of approximately a factor of two.

\begin{figure*}
 \includegraphics[height=5cm]{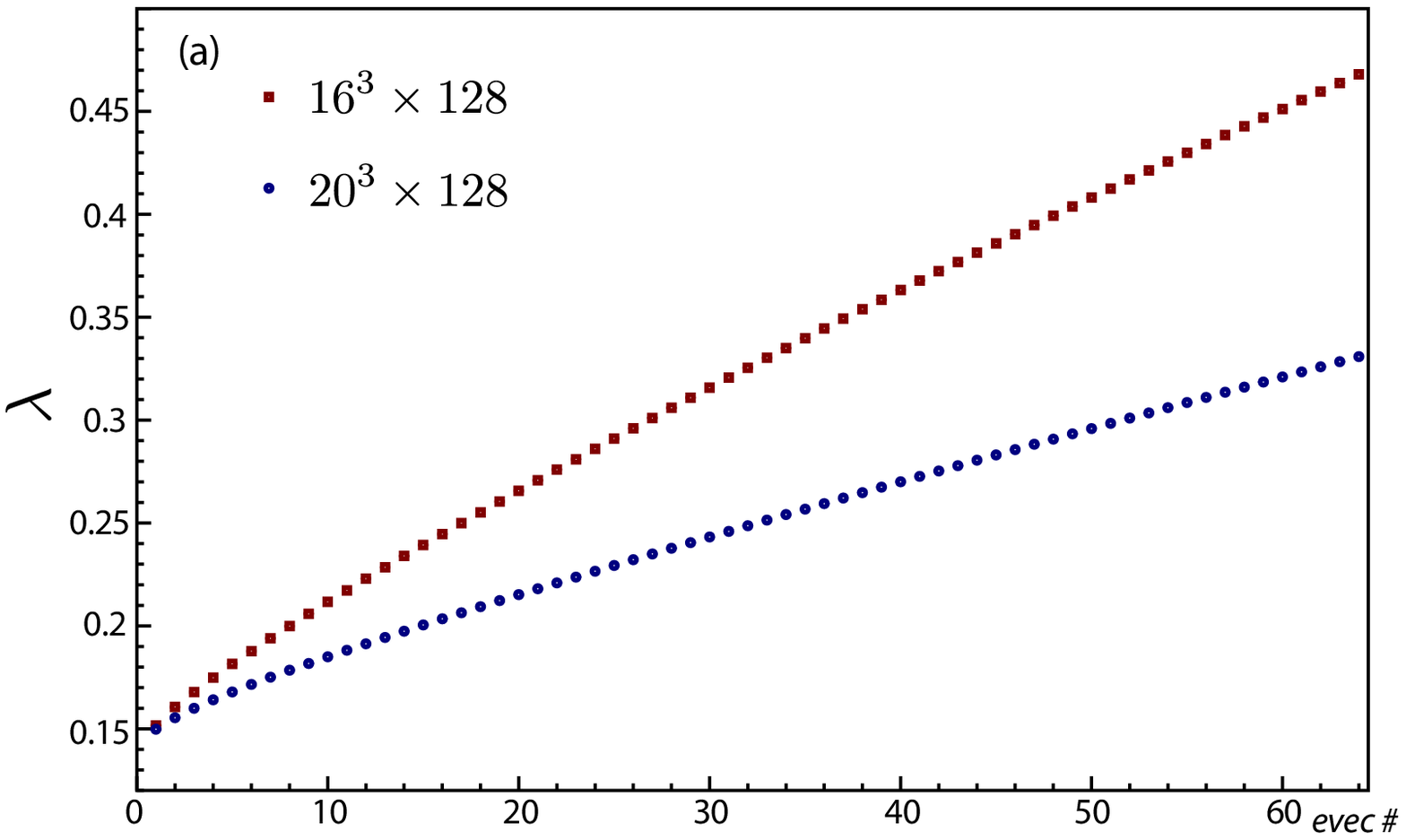}
\hspace{5mm}
\includegraphics[height=5cm]{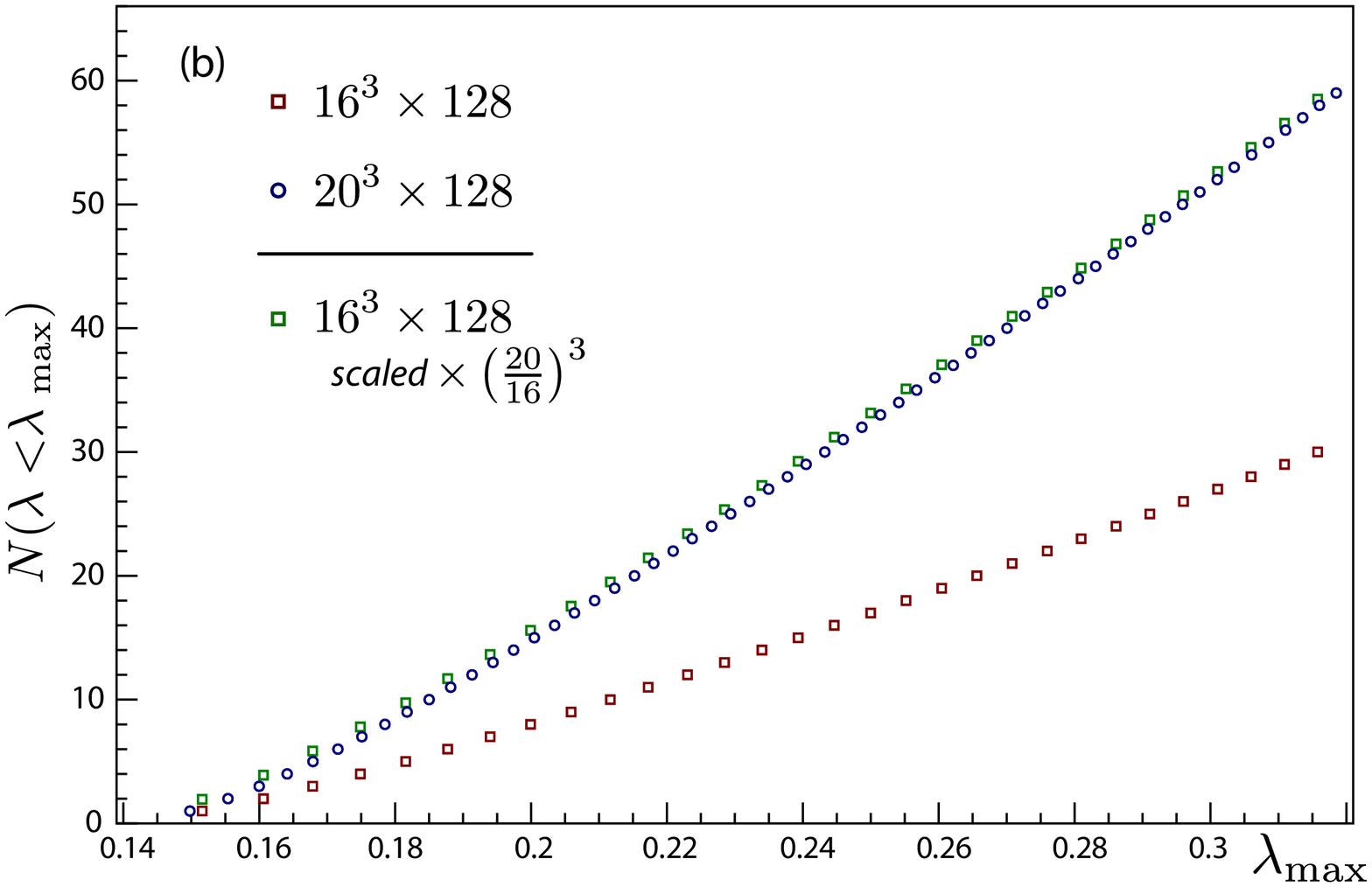}
\caption{\label{evals} (a) Eigenvalues of the laplacian on $16^3$ and $20^3$ lattices. (b) Number of eigenvectors with eigenvalue below $\lambda_{\mathrm{max}}$ on $16^3$ and $20^3$. Also shown is the $16^3$ data scaled by a factor of $\left(\frac{20}{16}\right)^3$.  }
\end{figure*}

A strikingly simple explanation for this effect comes through consideration of 
the eigenvalues belonging to these eigenvectors of the Laplacian operator, 
$\nabla^2 v^{(k)} = - \lambda_k v^{(k)}$. Fig.~\ref{evals}(a) shows the
eigenvalues measured on $16^3$ and $20^3$ lattices. One could consider choosing
the rank of the distillation operator by using all eigenvectors with
eigenvalue below a certain
cut-off. It is clear from the plot that on a larger volume this will require
more eigenvectors. This is displayed more directly in Fig.~\ref{evals}(b)
where the number of eigenvectors with eigenvalue less than a cut-off is
plotted. Approximately twice as many eigenvectors in
the $20^3$ case are required to have the same cut-off as in the $16^3$ case.
It appears from this analysis that the meson correlator signal is
effectively determined by the cut-off in eigenvalue used to construct the
distillation space.
This is equivalent to using a Heaviside approximation to smearing:
$J_{\lambda}(t) = \Theta( \lambda + \nabla^2(t)  ) = \sum_{k=1}^M \Theta(\lambda -  \lambda_k )  v^{(k)}(t) v^{(k)\dag}(t) = \sum_{k=1}^{N(\lambda)} v^{(k)}(t) v^{(k)\dag}(t).$

  \subsection{Baryon correlation functions}\label{subsec:results-baryons}
\begin{figure}
\includegraphics[width=0.9\textwidth]{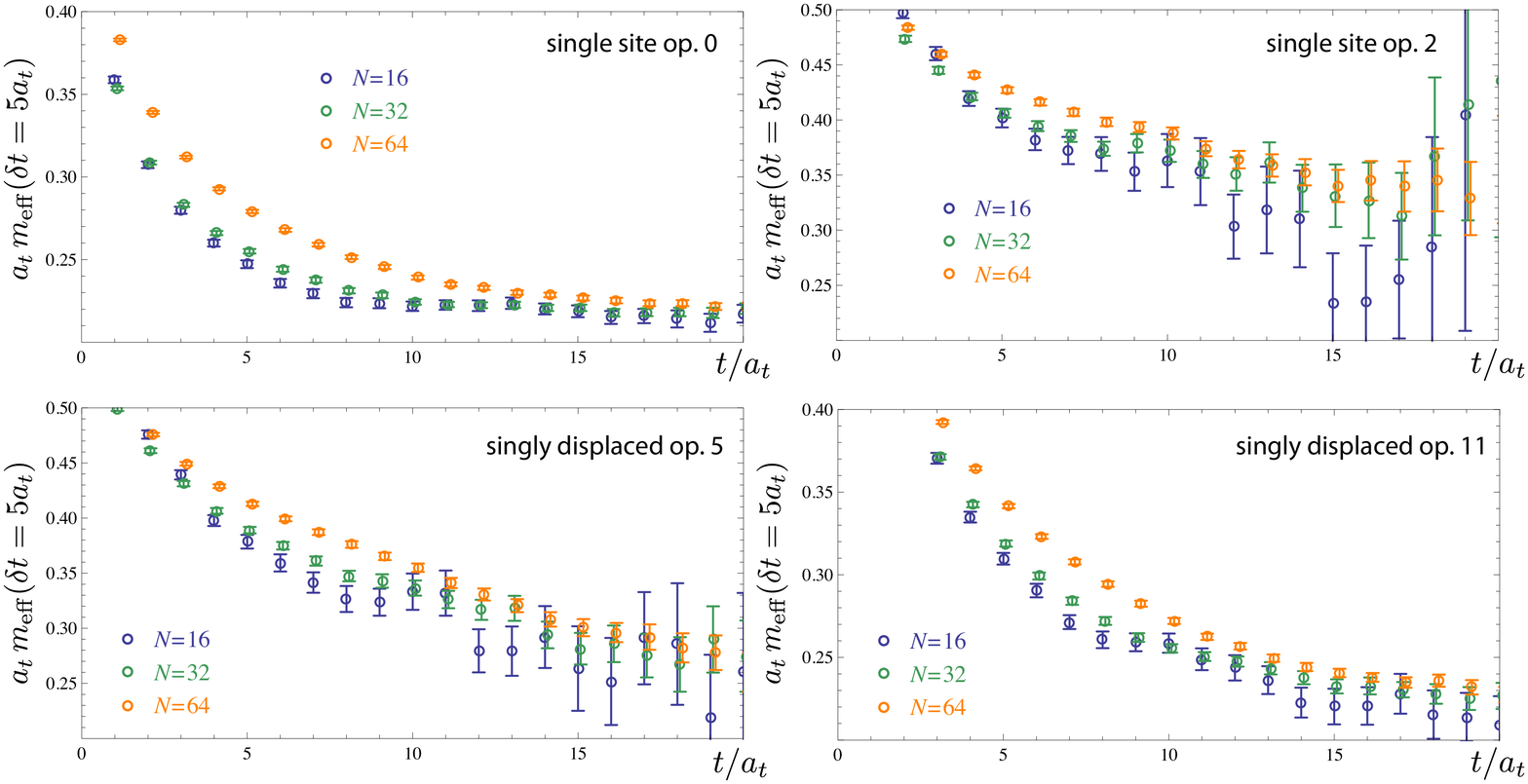}
\caption{\label{fig:baryon_meff_Nev}Effective-mass plot for two single-site and two singly displaced operator correlators in the nucleon $G_{1g}$ channel. 
}
\end{figure}

\begin{figure}
\includegraphics[width=0.5\textwidth]{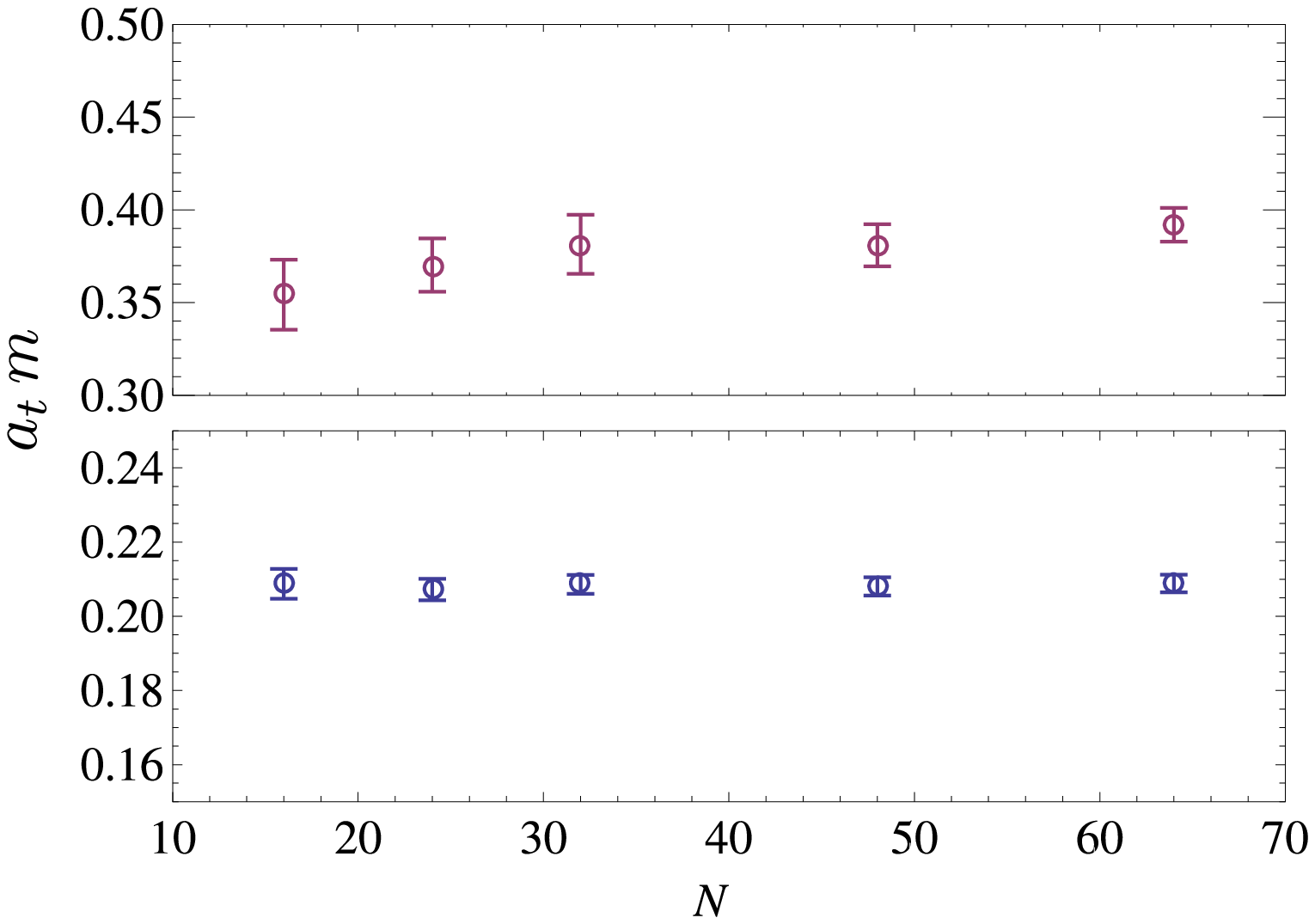}
\caption{\label{fig:baryon_Pmeff_Nev} The fitted energies for the first-excited (top) and ground state (bottom)  as functions of $N$.
}
\end{figure}

\begin{figure}
\includegraphics[width=0.9\textwidth]{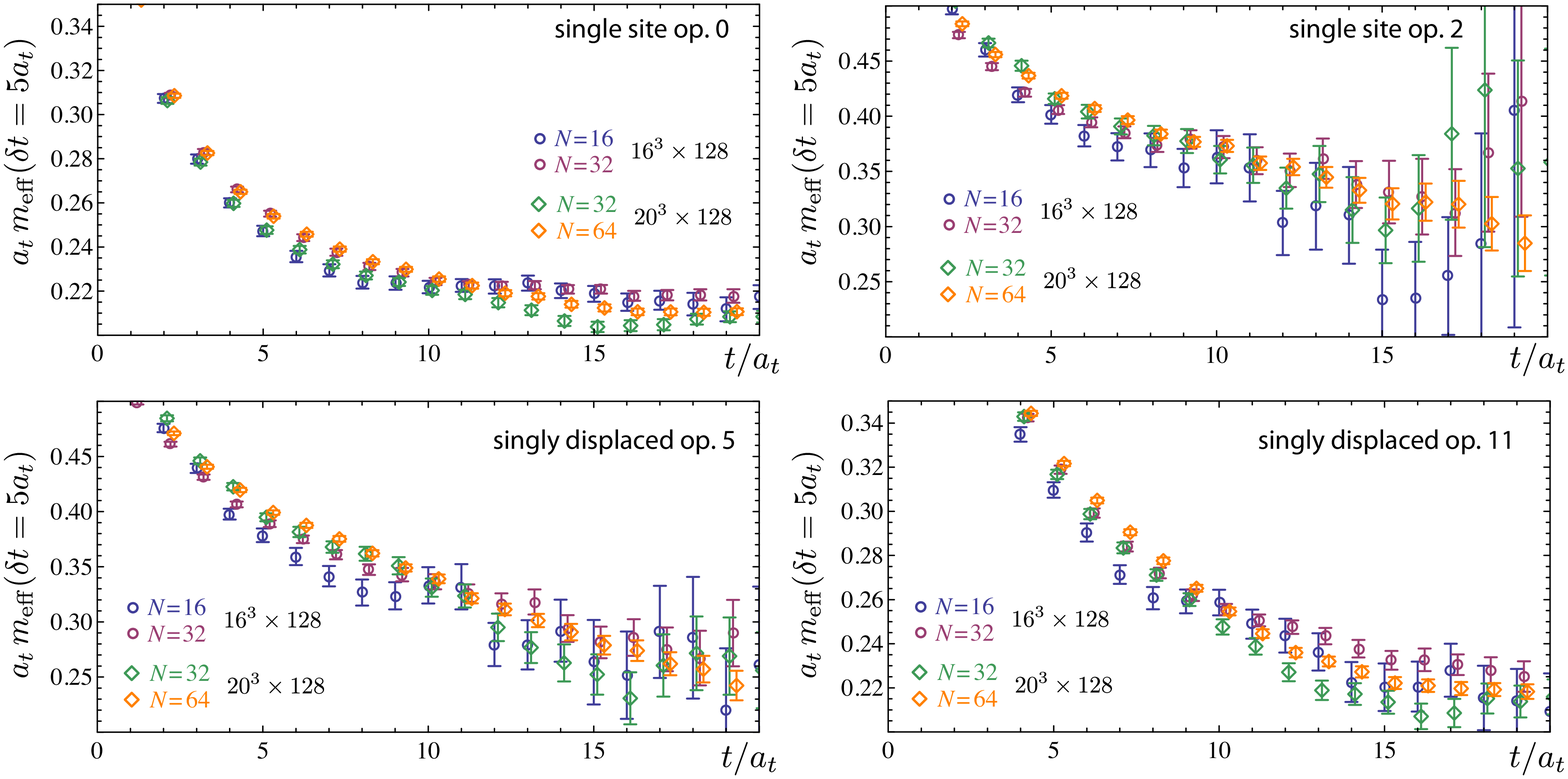}
\caption{\label{fig:baryon_volume_meff_Nev}Effective-mass plot for two
single-site and two singly displaced operator correlators from two volumes:
$16^3\times128$ and $20^3\times128$
}
\end{figure}

  Baryon correlation functions were evaluated using 
the displaced-quark operators described in
Refs.~\cite{Basak:2005aq,Basak:2005ir} and employed in spectrum studies
Refs.~\cite{Basak:2007kj, Bulava:2009jb}. 
The distillation method is demonstrated on four nucleon $G_{1g}$
operators, two of which are local to a single site with the remaining two
having a singly displaced quark field. The corresponding $4 \times 4$ matrix of correlators was computed on 316 configurations of the $16^3 \times 128$
lattice. Fig.~\ref{fig:baryon_meff_Nev} shows the effective-mass plots for
the diagonal correlators. A clear trend is observed, similar to that seen for 
the mesons, of the correlator having larger excited-state contribution but 
smaller statistical fluctuations as the number of eigenvectors is increased.
Modelling the correlator noise-to-signal ratio with Eq.~\ref{noise} 
gives best-fit exponents $p \sim 1.1(2)$, showing again that 
increasing the number of vectors decreases the noise considerably faster than 
simple statistical scaling. 
The variational method \cite{Michael:1985ne,Luscher:1990ck} was used to extract
the masses of the lowest two states in the $G_{1g}$ spectrum.
Fig.~\ref{fig:baryon_Pmeff_Nev} shows the dependence of the effective masses of
these states on the number of eigenvectors used to form the
distillation operator.  Consistent masses are found, with an increase in
statistical precision as the number of vectors is increased.
Fig.~\ref{fig:baryon_volume_meff_Nev} shows the diagonal effective masses
computed on $16^3$ and $20^3$ lattices. 
The similarity between the signals found before and after doubling
the number of vectors is not as obvious here as in the meson case.

  \subsection{Multi-meson correlation functions}\label{subsec:results-multi-mesons}
  One of the main motivations for defining distillation methods is to enable
efficient evaluation of correlation functions of creation operators which
resemble multi-hadron states where the component hadrons have definite
momentum. Reliable and efficient means of including these states in the operator
basis are important for addressing scattering and resonance properties.
The simplest case was tested, which comprises the
isospin-two two-pion state formed using Eq.~\ref{eqn:multi_meson_quark_op} with
the single particle operators $M(\vec{p}_\mathrm{rel})$ constructed using the
local operator $\bar{u}\gamma_5 d$. For this example, consider the
two-pion operator
\begin{equation}
  \chi_{MM}(|\vec{p}_\mathrm{rel}|^2, t) =
    \sum_{k\in {\cal R}(\vec{p}_\mathrm{rel})}
   \chi_M(k,t) \, \chi_M(-k,t)
\end{equation}
with total momentum zero, and the single-particle operators to have relative
momentum $\vec{p}_\mathrm{rel}$. The sum is over the set of momentum vectors
${\cal R}(\vec{p}_\mathrm{rel})$ related to $\vec{p}_\mathrm{rel}$ by lattice
rotations, to create a state that transforms trivially under rotations and so
the operator forms an $A_1$ irreducible representation of the cubic group.
In this test, all values of $|\vec{p}_\mathrm{rel}|$ between 0 and $2 \times
\frac{2\pi}{L}$ were considered, corresponding to $\vec{p}_\mathrm{rel} =
\frac{2\pi}{L}\vec{n}$ with $\vec{n} \in \{(0,0,0), (1,0,0), (1,1,0), (1,1,1),
  (2,0,0)\}$.

\begin{figure*}
 \includegraphics[width=10cm]{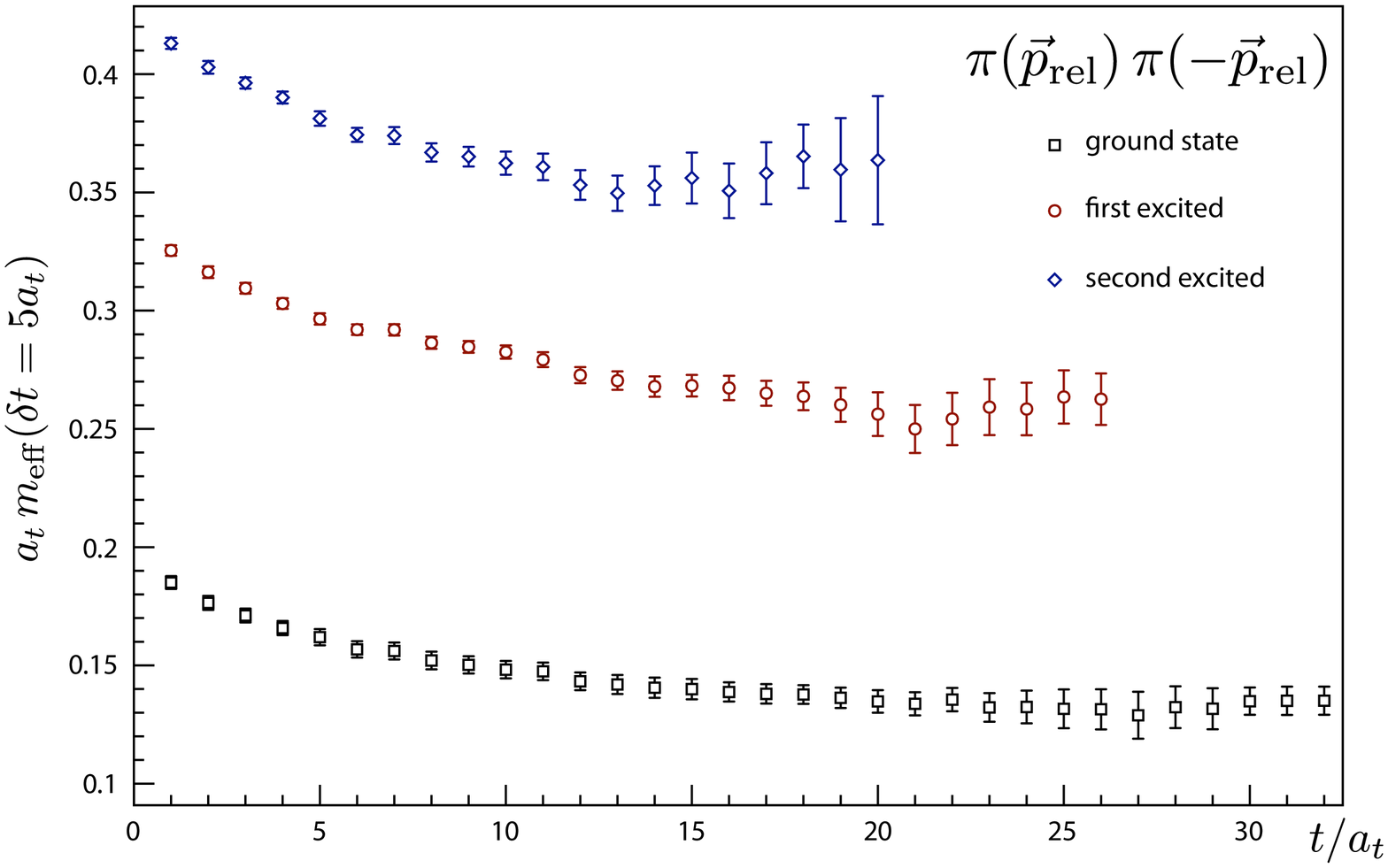}
\caption{\label{fig:twopi} Effective masses (five-timeslice shift) for the first three
principal correlators in the isospin-two $\pi(\vec{p}_\mathrm{rel})\pi(-\vec{p}_\mathrm{rel})$ channel ($A_1^{++}$) computed on
$16^3\times 128$ lattices. The $5\times 5$ correlation matrix
is constructed from $|\vec{p}_\mathrm{rel}|^2 \in [0,4]$.  Correlators constructed
through distillation with $N=64$ vectors on 100 configurations.}
\end{figure*}

A $5\times 5$ correlation matrix is formed between all of these two-pion 
operators which
is diagonalized by solving the generalized eigenvalue problem.  The
effective mass of the lowest three principal correlation functions measured on
100 configuration is shown in Fig.~\ref{fig:twopi}. Clear indication of
a signal for the second excited state is seen, even with this relatively
low level of statistics. Note that when using the conventional point-to-all 
method, the source operator cannot be projected onto definite momentum so the 
correlation matrix described above cannot be constructed in practice.

\section{Discussion and future directions \label{sec:discussion}}
  The results of Sec.~\ref{sec:results} demonstrate clearly that the light
hadrons can be studied using operators that excite only a very small number of 
extended degrees of freedom on each time-slice. The restricted set of modes 
needed to create a hadron means all elements of quark propagation needed to 
construct the correlation function can be measured directly. This does not 
preclude using stochastic methods \cite{Foley:2005ac} to reduce computational 
costs.

In the straightforward recipe described in this paper, the distillation space 
was constructed from the lowest eigenvectors of the lattice three-dimensional
Laplace operator on a time-slice. This definition is not unique and more
experimentation with different choices of distillation space might prove
useful. Also, the operator used in this work is diagonal in the quark spin 
index. Alternative choices of the distillation space might include non-trivial 
spin structure, using for example the vector space spanned by the lowest modes 
of the three-dimensional Dirac operator.

Recall that in the simplest implementation tested here when $N=M$,
the distillation operator becomes the identity and no smearing is performed. 
This is clearly an undesirable feature that is circumvented by the obvious 
variation on the method to include different weights for the eigenmodes. This 
more general distillation operator is then
\begin{equation}
        {\cal J}(t) = \sum_{k=1}^N v^{(k)}(t) \, f\big(\lambda_k(t)\big) 
              \, v^{(k)\dag}(t).
\end{equation}
Including weight functions on the eigenvectors can be used to suppress 
more rapidly fluctuating contributions when many eigenvectors are employed. 
Using exponential weights leads to increasingly accurate approximations to the 
conventional smearing algorithm of Eq.~\ref{eqn:smearing}. 

No increase in the number of eigenvectors would be expected as the continuum
limit is taken at a fixed physical volume since, as is seen in the data of
Sec.~\ref{subsec:results-wavefn}, the smooth distilled modes are
almost completely decoupled from the cut-off dynamics. This expectation is yet
to be tested numerically. 
One substantial possible drawback with the method is the need to increase the
rank of the distillation operator as the three-dimensional lattice volume
increases. The corresponding linear increase in the cost of solving the Dirac
equation leads to an algorithm for spectroscopy that scales like ${\cal
O}(V^2)$. 
It appears from the results we have presented that with the required
increase in rank with increasing volume one also obtains a decrease in 
statistical fluctuations.  The
possible solution to the issue of volume scaling may be to combine the method
with a suitably designed stochastic estimation scheme. This is under
investigation.

Note that the method does not give direct access to all elements of the quark
propagator, only those relevant to low-energy spectroscopy. This would be a
limitation in calculations involving isoscalar operator insertions, where
disconnected loops with all momentum components of the quark field are needed.
A well-known example of such a computation would be the evaluation of the 
strangeness content of the nucleon.

\section{Summary}
  In this paper, a simple new quark-field smearing has been proposed and tested in
numerical investigations using QCD gauge fields with $N_f=2+1$ dynamical
flavors. There is substantial freedom in the construction of creation operators
for hadrons.  This has been exploited in this work to define an algorithm 
that applies a low-rank projection operator to the quark fields of the path 
integral on each time-slice in order to define smooth modes for subsequent 
operator construction.  This makes affordable the evaluation of all 
information about quark propagation required to measure distilled hadronic 
correlation functions.  Once the propagation matrix has been evaluated, 
correlations between arbitrary choices of creation and annihilation 
operators can be determined without requiring any further inversions. The 
theoretical framework needed to extend the scope of measurements using this 
technology to include matrix elements was described briefly. 

Some initial tests of the method demonstrate that restricting quark fields to
lie in a very low-dimensional space of smooth fields does not substantially 
degrade the quality of Monte Carlo evaluations of correlation functions and 
in 
many cases, statistical accuracy is enhanced. The most substantial drawback, 
still to be overcome, is the growth in cost of the algorithm as the lattice 
volume in physical units increases. Since the method is concerned with 
long-distance physics, no extra difficulties in approaching the continuum 
limit are anticipated, but this would need to be confirmed by practical 
testing. 

The method is currently in its infancy and a number of research directions are
outlined in this paper. The primary focus of the development efforts of this
collaboration is to include stochastic evaluation of correlation 
functions in an optimal way and results will appear soon. The collaboration 
will use the techniques outlined in this paper in a range of spectroscopy 
determinations in the near future.

\section*{Acknowledgements}
  The Chroma software suite~\cite{Edwards:2004sx} was used to perform this work 
on clusters at Jefferson Laboratory using time awarded under the USQCD 
Initiative.

We thank Sin\'ead Ryan for many helpful comments on this manuscript. 
MP is supported by Science Foundation Ireland under research grant
07/RFP/PHYF168.  MP is extremely grateful for the generous hospitality of the 
theory center at TJNAF during the early stages of this work. 
JB, JF and CM are supported by National Science Foundation grant numbers 
NSF-PHY-0653315 and NSF-PHY-0510020. 
KJJ is supported by grant number NSF-PHY-0704171. 

Authored by Jefferson Science Associates, LLC under U.S. DOE Contract No.
DE-AC05-06OR23177. The U.S. Government retains a non-exclusive, paid-up,
irrevocable, world-wide license to publish or reproduce this manuscript for
U.S. Government purposes.

\bibliography{distillation}

\end{document}